\documentclass[11pt]{article}

\usepackage{fullpage}
\usepackage{amsmath}

\usepackage{pst-all}

\usepackage{latexsym}
\usepackage{amsmath}
\usepackage{amstext}
\usepackage{amsfonts}
\usepackage{amsopn}

\usepackage{ifthen}

\newtheorem{theorem}{Theorem}[section]
\newtheorem{lemma}[theorem]{Lemma}
\newtheorem{corollary}[theorem]{Corollary}
\newtheorem{proposition}[theorem]{Proposition}
\newtheorem{definition}[theorem]{Definition}

\newenvironment{ProofDummyEnv}{}{}

\newenvironment{proof}[1][none]{\begin{proofby}[#1]{}}{\end{proofby}}

\newenvironment{proofby}[2][none]{\par\noindent{\bf Proof:} #2
\renewenvironment{ProofDummyEnv}{\begin{#1}}{\end{#1}}%
\begin{ProofDummyEnv}}%
{\QED\end{ProofDummyEnv}}

\newcommand{\QED}{\nopagebreak\hfill $\Box$}

\newcommand{\prob}[2][]{\text{\bf Pr}\ifthenelse{\not\equal{}{#1}}{_{#1}}{}\!\left[#2\right]}
\newcommand{\expect}[2][]{\text{\bf E}\ifthenelse{\not\equal{}{#1}}{_{#1}}{}\!\left[#2\right]}

\newcommand{\setsize}[1]{\left| #1 \right|}

\newcommand{\suchthat}{\ :\ }

\DeclareMathOperator{\argmax}{argmax}

\title{Optimal Mechanism Design and Money Burning}

\author{Jason D. Hartline\thanks{Electrical Engineering and Computer
Science, Northwestern University, Evanston, IL.  This work was done
while author was at Microsoft Research, Silicon Valley.
Email: {\tt hartline@eecs.northwestern.edu}.} \and
Tim Roughgarden\thanks{Department of Computer Science, Stanford
University, 462 Gates Building, 353 Serra Mall, Stanford, CA 94305.
Supported in part by NSF CAREER Award CCF-0448664, an ONR Young
Investigator Award, and an Alfred P. Sloan Fellowship.  Email: {\tt
tim@cs.stanford.edu}.}}

\date{First Draft: January 2007; this draft: April 2008}

\newcommand{\Xcomment}[1]{}

\DeclareMathOperator{\RSOL}{RSOL}

\newcommand{\bid}{b}

\newcommand{\bidi}[1][i]{{\bid_{#1}}}

\newcommand{\val}{v}
\newcommand{\vals}{{\mathbf \val}}
\newcommand{\valsmi}{{\mathbf \val}_{-i}}
\newcommand{\vali}[1][i]{{\val_{#1}}}
\newcommand{\valith}[1][i]{{\val_{(#1)}}}

\newcommand{\util}{u}

\newcommand{\utili}[1][i]{{\util_{#1}}}

\newcommand{\virt}{\varphi}

\newcommand{\ironvirt}{\bar{\virt}}

\newcommand{\dist}{F}
\newcommand{\dists}{{\mathbf \dist}}
\newcommand{\disti}[1][i]{{\dist_{#1}}}

\newcommand{\dens}{f}

\newcommand{\price}{p}
\newcommand{\prices}{{\mathbf \price}}
\newcommand{\pricei}[1][i]{{\price_{#1}}}

\newcommand{\alloc}{x}
\newcommand{\allocs}{{\mathbf \alloc}}

\newcommand{\alloci}[1][i]{{\alloc_{#1}}}

\newcommand{\feasibles}{{\cal X}}

\newcommand{\expval}{\mu}

\newcommand{\marg}{\vartheta}

\newcommand{\ironmarg}{\bar{\marg}}

\newcommand{\ironmargi}[1][i]{{\ironmarg_{#1}}}

\newcommand{\cost}{c}

\newcommand{\bm}{{\cal G}}
\DeclareMathOperator{\Mye}{Opt}
\DeclareMathOperator{\Lottery}{Lot}

\begin{document}

\maketitle

\begin{abstract}
Mechanism design is now a standard tool in computer science for
aligning the incentives of self-interested agents with the objectives
of a system designer.  
There is, however, a fundamental disconnect between the traditional
application domains of mechanism design (such as auctions) and those arising
in computer science (such as networks): while monetary {\em transfers}
(i.e., payments) are essential for most of the known positive results
in mechanism design, they are undesirable or even technologically
infeasible in many computer systems.
Classical impossibility results imply
that the reach of mechanisms without transfers is severely limited.

Computer systems typically do have the ability to
reduce service quality---routing systems can drop or delay traffic,
scheduling protocols can delay the release of jobs,
and computational payment schemes can require
computational payments from users (e.g., in spam-fighting
systems).
Service degradation is tantamount to requiring that users {\em burn money},
and such ``payments'' can be used to influence the preferences of the
agents at a cost of degrading the social surplus.

We develop a framework for the design and analysis of {\em
money-burning mechanisms} to maximize the residual surplus---the
total value of the chosen outcome minus the payments required.  Our
primary contributions are the following.
\begin{itemize}

\item We define a general template for prior-free optimal
mechanism design that explicitly connects Bayesian
optimal mechanism design, the dominant paradigm in economics, with
worst-case analysis.
In particular, we establish a general and principled way to
identify appropriate performance benchmarks for prior-free optimal mechanism
design.

\item For general single-parameter agent settings,
we characterize the Bayesian optimal money-burning mechanism.  

\item For multi-unit auctions, we design a near-optimal prior-free
money-burning mechanism: for every valuation profile, its expected
residual surplus is within a constant factor of our benchmark, the
residual surplus of the best Bayesian optimal mechanism for this profile.

\item For multi-unit auctions, we quantify the benefit of
general transfers over money-burning: optimal money-burning mechanisms
always obtain a logarithmic fraction of the full social surplus, and
this bound is tight.

\end{itemize}
\end{abstract}

\newpage

\section{Introduction}
Mechanism design is now a standard tool in computer science for
designing resource allocation protocols (a.k.a.\ mechanisms) in
computer systems used by agents with diverse and selfish interests.
The goal of mechanism design is to achieve non-trivial optimization
even when the underlying data---the preferences of participants---are
unknown a priori.
Fundamental for most positive results in mechanism design are monetary
transfers (i.e., payments) between participants.  For example, in the
surplus-maximizing VCG mechanism~\cite{vic-61,cla-71,gro-73}, such
transfers enable the mechanism designer to align fully  the incentives
of the agents with the system's objective.

Most computer systems differ from classical environments for
mechanism design, such as traditional markets and auctions, in that
monetary transfers are unpopular, undesirable, or technologically
infeasible.  
It is sometimes possible to design mechanisms that eschew transfers
completely; see~\cite{SV07} for 
classical results in economics and~\cite{FSS07,LSZ08} for
recent applications in interdomain routing.  Unfortunately, negative
results derived from Arrow's Theorem~\cite{arr-51,gib-73,sat-75} imply
that the reach of mechanisms without transfers is severely limited.

The following observation motivates our work: 
{\em computer systems
typically have the ability to arbitrarily reduce service quality}.
For example, routing systems can drop or delay traffic
(e.g.~\cite{CDR06}), scheduling protocols can 
delay the release of jobs (e.g.~\cite{CKN04}), and computational
payment schemes allow a mechanism to demand 
computational payments from agents (e.g., in spam-fighting
systems~\cite{DN-92,DGN-03,LC-06}).\footnote{Computational payment schemes
do not need the infrastructure required by micropayment schemes.
One can interpret our results as analyzing the power of computational
payments, which were first devised for spam-fighting, in a general
mechanism design setting.}
Such service degradation can 
be used to align the preferences of the agents with the social
objective, at a cost: {\em these ``payments'' also degrade the social
surplus.}  

We develop a framework for the design and analysis of {\em
money-burning mechanisms}---mechanisms that can employ arbitrary
payments and seek to maximize the {\em residual surplus}, defined as
the total value to the participants of the chosen outcome minus the
sum of the (``burnt'') payments.\footnote{We assume that valuations
and burnt payments are measured in the same units.  In other words,
there is a known mapping between decreased service quality (e.g.,
additional delay) and lost value (e.g., dollars).  This mapping can be
different for different participants, but it must be publicly known
and map onto $[0,\infty)$.  See Section~\ref{sec:conc} for further
discussion.}  Such mechanisms must trade off the social cost of
imposing payments with the ability to elicit private information from
participants and thereby enable accurate surplus-maximization.  For
example, suppose we intend to award one of two participants access to
a network.  Assume that the two agents have valuations (i.e., maximum
willingness to pay)~$\vali[1]$ and~$\vali[2] \le \vali[1]$ for
acquiring access, and that these valuations are private (i.e., unknown
to the mechanism designer).
The {\em Vickrey} or {\em second-price auction}~\cite{vic-61} would
award access to agent~1, charge a payment of $\vali[2]$, and thereby
obtain residual surplus $\vali[1] - \vali[2]$.  
A {\em lottery} would award access to
an agent chosen at random, charge nothing, and achieve a
(residual) surplus of $(\vali[1]+\vali[2])/2$, a better result if and
only if $\vali[1] < 3\vali[2]$.  Even in this trivial scenario, it is not clear
how to define (let alone design) an optimal money-burning
mechanism.\footnote{Indeed, it follows from our results that in some
settings lotteries are optimal (i.e., money-burning is useless); in
others, Vickrey auctions are optimal; and sometimes, neither is
optimal.}

Our goal is to rigorously answer the following two questions:
\begin{enumerate}

\item 
What is the optimal money-burning mechanism?

\item 
How much more powerful are mechanisms with monetary transfers than
money-burning mechanisms?

\end{enumerate}

\paragraph{Our Results.}
Our first contribution is to identify a general template for
prior-free (i.e., worst-case) optimal mechanism design.
The basic idea is to characterize the set of 
mechanisms that are Bayesian optimal for some i.i.d.\ distribution on
valuations, and then define a prior-free performance benchmark that
corresponds to competing simultaneously with all of these on a fixed
(worst-case) valuation profile.
The template, which we detail below, is general and we expect it to
apply in many mechanism design settings beyond money-burning
mechanisms.

Second, we characterize Bayesian optimal money-burning
mechanisms---the incentive-compatible mechanisms with maximum-possible
expected residual surplus.
Our characterization applies to general single-parameter
agents, meaning that the preferences of each agent is naturally
summarized by a single real-valued valuation, with independent but
not necessarily identically distributed valuations.
The characterization unifies
results in the economics literature~\cite{CK06,MM} and also extends
them in two important directions.  First, the results
in~\cite{CK06,MM} concern only multi-unit auctions, where $k$
identical units of an item can be allocated to agents who each desire
at most one unit.  Our characterization applies to the general,
possibly asymmetric, setting of single-parameter agents; for example,
agents could be seeking disjoint paths in a multicommodity
network.\footnote{Multi-unit auctions model symmetric situations, as
when each agent seeks a path from a common source~$s$ to a common
destination~$t$; here, the number~$k$ of units equals the number of
edges in a minimum $s$-$t$ cut.}  In addition, for multi-unit
auctions, we give a simple description of the optimal mechanism even
when the ``hazard rate'' of the valuation distribution is not monotone
in either direction.  
This important case is the most technically interesting and
challenging one, and it has not been considered in detail in the
literature.

Third, for multi-unit auctions, we 
design a mechanism that is approximately optimal in the worst case.  
We derive our benchmark using our characterization of Bayesian optimal
mechanisms restricted to i.i.d.\ valuations and symmetric mechanisms.
We prove that such mechanisms are always well approximated by a
{\em $k$-unit $p$-lottery}, defined as follows: order the agents
randomly, sequentially make each agent a take-it-or-leave-it offer of
$p$, and stop after either~$k$ items have been allocated or all agents
have been considered.  This result reduces the design of a
constant-approximation prior-free money-burning mechanism to the
problem of approximating the residual surplus achieved by the optimal
$k$-unit $p$-lottery.  Our prior-free mechanism obtains a constant
approximation of this benchmark using random sampling to select a good
value of $p$.  Surprisingly, we accomplish this even when $k$ is very
small (e.g., $k=1$).\footnote{Previous experience with
prior-free mechanism design, e.g., for digital goods, suggests that
conditions like ``two or more winners'' might be necessary to achieve
a constant approximation.  See Section~\ref{sec:conc}
for further discussion.}
Our benchmark definition ensures that such a guarantee is strong: for
example, if valuations are drawn from some unknown
i.i.d.~distribution~$\dists$, our mechanism obtains a constant
fraction of the expected residual surplus of an optimal mechanism
tailored specifically for~$\dists$.

Finally, for multi-unit auctions, we provide a price-of-anarchy-type
analysis that measures the social cost of burnt payments.  
Recall that the full surplus is achievable with monetary
transfers using the Vickrey-Clarke-Groves (VCG) mechanism.
We prove that the largest-possible relative loss in surplus due to
money-burning is precisely logarithmic in the number of participants,
in both the Bayesian and worst-case settings.
Indeed, our near-optimal money-burning mechanism always obtains
residual surplus within a logarithmic factor of the full surplus.  This
result suggests that the cost of implementing money-burning (e.g.,
computational payments) rather than general transfers (e.g.,
micropayments) in a system is relatively modest.  Further, our
positive result contrasts with the linear lower bound that we
prove on the fraction of the full surplus obtainable by mechanisms
without any kind of payments.

\paragraph{A Template for Prior-Free Auction Design.}
The following template forges an explicit connection between the
Bayesian analysis of Bayesian optimal mechanism design, the dominant
approach in economics, and the worst-case analysis of prior-free optimal
mechanism design, the ubiquitous approach in theoretical computer science.
Its goal is to fill a fundamental gap in prior-free optimal mechanism
design methodology: the selection of an appropriate performance
benchmark.
\begin{enumerate}

\item \label{step:bayesian} Characterize the Bayesian optimal
mechanism for every i.i.d.~valuation distribution.

\item \label{step:benchmark} Interpret the behavior of the symmetric,
ex post incentive compatible, Bayesian optimal mechanism 
for every i.i.d.\ distribution on an
arbitrary valuation profile 
to give a distribution-independent {\em benchmark}.

\item \label{step:design} Design a single ex post incentive compatible
mechanism that approximates the above benchmark on every
valuation profile; the performance ratio of such a mechanism 
provides an upper bound on that of the optimal prior-free mechanism.

\item \label{step:lowerbound} Obtain lower bounds on the best
performance ratio possible in this framework
by exhibiting a distribution over valuations such that the
ratio between the expected value of the benchmark and the performance
of the Bayesian optimal mechanism for the given distribution is large.

\end{enumerate}
In hindsight, this approach has been employed implicitly in the context of
(profit-maximizing) digital good auctions \cite{GHW-01,GHKSW-06}.
However, the simplicity 
of the digital good auction problem obscures the importance of the first
two steps, as the Bayesian optimal digital good auction is trivial:
offer a posted price.  For money-burning mechanisms, the benchmark we
identify in Step~\ref{step:benchmark} is not a priori obvious.

\paragraph{Further Related Work.} 
McAfee and McMillan study collusion among bidders in multi-unit
auctions \cite{MM}.  In a {\em weak cartel}, where the agents wish to
maximize the cartel's total utility but are not able to make side
payments amongst themselves, payments made to the auctioneer are
effectively burnt.  The optimization and incentive problem faced by
the grand coalition in a multi-unit auction is similar to the
auctioneer's problem in our money-burning setting; therefore, results
for weak cartels follow from similar analyses to ours \cite{MM,C07}.

Our characterization of Bayesian optimal money-burning mechanisms
builds on analysis tools developed for profit maximization in Bayesian
settings (see Myerson~\cite{mye-81} and Riley and
Samuelson~\cite{RS-81}) that apply in general single parameter
settings (see, e.g., \cite{HK-07}).  Independently from our work,
Chakravarty and Kaplan~\cite{CK06} describe the optimal Bayesian
auction in multi-unit money-burning settings.  Our work extends this
analysis to general single-parameter agent settings with explicit
focus on the case where the hazard rate is not monotone in either
direction.  Our paper is the first to study the relative power of
money-burning mechanisms and mechanisms with or without transfers.  It
is also the first to consider prior-free money-burning mechanisms.

Our results that quantify the benefit of transfers have analogs in
the price of anarchy literature, specifically in the standard
(nonatomic) model of selfish routing (e.g.~\cite{book}).  Namely, full
efficiency is achievable in this model with general transfers, in the
form of ``congestion prices''; without transfers the outcome is a Nash
equilibrium, with inefficiency measured by the price of anarchy; and
with burnt transfers (``speed bumps'' or other artificial delays) it
is generally possible to recover some but not all of the efficiency
loss at equilibrium~\cite{CDR06}.  

There are several other studies that view transfers to an auctioneer
as undesirable; however, these works are technically unrelated to ours.
We already noted recent work on incentive-compatible interdomain
routing without payments~\cite{FSS07,LSZ08}.
Moulin~\cite{mou-06} and Guo and Conitzer~\cite{GC-07} independently
studied how to redistribute the payments of the VCG mechanism 
in a multi-unit auction among the participants (using general
transfers) to minimize the total payment to the auctioneer.
Finally, as already mentioned, our prior-free techniques are related
to recent work on profit maximization
(e.g.,~\cite{GHW-01,FGHK-02,BBHM-05}) and there is a related
literature on the problem of cost minimization, a.k.a.~frugality 
(e.g.,~\cite{AT-02,tal-03,ESS-04,KKT-05}).

\section{Bayesian Optimal Money Burning}\label{sec:bayesian}

In this section we study optimal
money-burning mechanism design from a standard economics viewpoint,
where agent valuations are drawn from a known {\em prior distribution}.
This will complete the first step of our template for
prior-free optimal mechanism design.

\paragraph*{Mechanism design basics.}
We consider mechanisms that provide a good or service to a subset of $n$
agents.  The outcome of such a mechanism is an {\em allocation vector},
$\allocs = (\alloci[1],\ldots,\alloci[n])$, where $\alloci$ is~1 if
agent~$i$ is served and~0 otherwise, and a {\em payment vector}, $\prices
= (\pricei[1],\ldots,\pricei[n])$.  In this paper, the payment
$\pricei$ is the amount of money that agent~$i$ must ``burn''.  We
allow the set of feasible allocation vectors, $\feasibles$, to be
constrained arbitrarily; for example, in a multi-unit auction with~$k$
identical units of an item, the feasible allocation vectors are those
$\allocs \in \feasibles$ with $\sum_i \alloci \leq k$.

We assume that each agent~$i$ is {\em risk-neutral}, has a privately
known valuation $\vali$ for receiving service, and aims to maximize
their (quasi-linear) utility, defined as $\utili = \vali \alloci -
\pricei$.  We denote the {\em valuation profile} by $\vals =
(\vali[1],\ldots,\vali[n])$.

Our mechanism design objective is to maximize the {\em residual
surplus}, defined as
$$
\sum\nolimits_i (\vali\alloci - \pricei)
$$ 
for a valuation profile $\vals$, a feasible allocation $\allocs$, and
payments $\prices$.
If the payments were transferred to the seller then the resulting
{\em social surplus}
would be $\sum_i \vali\alloci$; however, in
our setting the payments are burnt and the social surplus is equal to
the residual surplus.  

\paragraph*{Bayesian mechanism design basics.}
In this section,
we assume that the agent valuations are drawn
{i.i.d.}~from a publicly known distribution with cumulative
distribution function $\dist(z)$ and probability density
function $\dens(z)$.  
We let $\dists$ denote the joint (product) distribution of agent
values.
See Section~\ref{sec:conc} for a generalization to general product
distributions.

We consider the problem of implementation in Bayes-Nash
equilibrium.  Agent $i$'s strategy is a mapping from their private
value $\vali$ to a course of actions in the mechanism.  
The distribution on valuations $\dists$ and a strategy profile induce a
distribution on agent actions.  These agent actions are in {\em Bayes-Nash
equilibrium} if no agent, given their own valuation and the
distribution on other agents' actions, can improve its expected payoff
via alternative actions.
By the {\em revelation principle}~\cite{mye-81}, we can restrict our
attention to single round, sealed bid, {\em direct} mechanisms in
which {\em truthtelling}, i.e., submitting a bid $\bidi$ equal to the
private value $\vali$, is a Bayes-Nash equilibrium.  It will turn out
that there is always an optimal mechanism that is not only Bayesian
incentive compatible but also {\em dominant strategy incentive
compatible}, meaning truthtelling is an optimal agent strategy for every
strategy profile of the other agents.

An {\em allocation rule}, $\allocs(\vals)$, is the mapping 
(in the truthtelling equilibrium)
from agent valuations to the outcome of the mechanism.
Similarly the {\em payment rule}, $\prices(\vals)$, is the mapping
from valuations to payments.  Given an allocation rule $\allocs(\vals)$,
let $\alloci(\vali)$ be the probability that agent $i$ is allocated
when its valuation is $\vali$ (over the probability distribution on
the other agents' valuations): $\alloci(\vali) =
\expect[\valsmi]{\alloci(\vali,\valsmi)}.$ 
Similarly define $\pricei(\vali)$.  
Positive transfers from the mechanism to the agents
are not allowed and we require ex interim individual rationality 
(i.e., that
non-participation in the mechanism is an allowable agent strategy).
The following lemma is the standard characterization of the allocation
rules implementable by Bayesian incentive-compatible mechanisms and
the accompanying (uniquely defined) payment rule.
\begin{lemma}\cite{mye-81}
\label{l:bic}\label{l:ic} 
Every Bayesian incentive compatible mechanism satisfies:
\begin{enumerate}
\item Allocation monotonicity: for all $i$ and $\vali > \vali'$, 
$\alloci(\vali) \geq \alloci(\vali').$

\item Payment identity: for all $i$ and $\vali$, 
$\pricei(\vali) = \vali\alloci(\vali) - \int_0^{\vali}
  \alloci(\val) d\val.$
\end{enumerate}
\end{lemma}

\paragraph*{Virtual valuations.}
Assume for simplicity that the distribution $\dist$ has support
$[a,b]$ and positive density throughout this interval.
Myerson~\cite{mye-81} defined ``virtual
valuations'' and showed that they characterize the expected payment
of an agent in a Bayesian incentive compatible mechanism.

\begin{definition}[virtual valuation for payment~\cite{mye-81}]
If agent $i$'s valuation is distributed according to
  $\dist$, then its {\em virtual valuation for payment} is
$$
\virt(\vali) = \vali - \tfrac{1-\dist(\vali)}{\dens(\vali)}.
$$
\end{definition}

\begin{lemma}\cite{mye-81} 
In a Bayesian incentive-compatible mechanism with allocation rule
$\allocs(\cdot)$,
the expected payment of agent~$i$ satisfies
$$
\expect[\vals] {\pricei(\vals)} = \expect[\vals] {\virt(\vali)\alloci(\vals)}.
$$
\end{lemma}

Myerson uses this correspondence to design optimal
mechanisms for profit-maximization.  The optimal mechanism 
for a given distribution is the one that maximizes the {\em virtual
surplus} (for payment).

\begin{definition}[virtual surplus]
For virtual valuation function $\virt(\cdot)$ and valuations $\vals$,
the {\em virtual surplus} of allocation $\allocs$ is
$$
\sum\nolimits_i \virt(\vali) \alloci.
$$
\end{definition}

Our objective is to maximize the residual surplus, $\sum_i (\vali
\alloci(\vals) - \pricei(\vals))$, which we can do quite easily using
virtual valuations.  To justify our terminology, below, notice that an
agent's utility is $\utili(\vals) = \vali \alloci(\vals) -
\pricei(\vals)$, and our objective of residual surplus maximization is
simply that of maximizing the expected utility of the agents,
$\expect[\vals]{\sum_i \utili(\vals)}$.  We define a {\em virtual
valuation for utility} by simply plugging in the virtual valuation for
payments into the equation that defines utility.

\begin{definition}[virtual valuation for utility] 
If agent $i$'s valuation is distributed according to
  $\dist$, then its {\em virtual valuation for utility} is
$$
\marg(\vali) = \tfrac{1 - \dist(\vali)}{\dens(\vali)}.
$$
\end{definition}
This quantity is also known as the ``information rent'' or ``inverse
hazard rate function''.  
Treating it as a virtual valuation of sorts,
we can generalize the theory of optimization by virtual valuations,
beginning with the following lemma.

\begin{lemma} 
\label{l:vv}
In a Bayesian incentive-compatible mechanism with allocation rule
$\allocs$, the expected utility of agent $i$ satisfies
$$
\expect[\vals] {\utili(\vals)} = \expect[\vals] {\marg(\vali)\alloci(\vals)}.
$$
\end{lemma}

We can conclude from this that the Bayesian optimal mechanisms for
residual surplus are precisely those that maximize the expected virtual
surplus (for utility) subject to feasibility and monotonicity of the
allocation rule.  In other words, 
we should choose a feasible allocation vector
$\allocs(\vals)$ to maximize $\sum_i \marg(\vali) \alloci(\vals)$
for each $\vals$,
subject to monotonicity of $\alloci(\vali)$.  It is easy to see that if
$\marg(\cdot)$ is monotone non-decreasing in $\vali$, then choosing
$$
\allocs(\vals) \in \argmax_{\allocs' \in \feasibles} \sum\nolimits_i \marg(\vali)\alloci'
$$
results in a monotone allocation rule.
Unfortunately $\marg(\cdot)$ is often not monotone non-decreasing;
indeed, under the standard ``monotone hazard rate'' assumption,
discussed further below, $\marg(\cdot)$ is monotone {\em in the
wrong direction}.

\paragraph*{Ironing.}
We next generalize an ``ironing'' procedure of Myerson~\cite{mye-81} that transforms a possibly non-monotone virtual
valuation function into an
{\em ironed virtual valuation} function that is monotone;
the optimization approach of the previous paragraph can then be
applied to these ironed functions to obtain a monotone allocation 
rule.  Further, the ironing procedure preserves the target objective,
so that an optimal allocation for the ironed virtual valuations is
equal to the optimal monotone allocation for 
the original virtual valuations.

\begin{definition}[ironed virtual valuations~\cite{mye-81}]
\label{d:ironing}
Given a distribution function $\dist(\cdot)$ with virtual valuation
(for utility) function $\marg(\cdot)$, the {\em ironed virtual
valuation function}, $\ironmarg(\cdot)$, is constructed as follows:
\begin{enumerate}
\item For $q \in [0,1]$, define $h(q) = \marg(\dist^{-1}(q))$.
\item Define $H(q) = \int_0^q h(r) dr$.
\item Define $G$ as the convex hull of~$H$ --- the largest
convex function bounded above by~$H$ for all $q \in [0,1]$.
\item Define $g(q)$ as the derivative of $G(q)$, where defined, and
  extend to all of $[0,1]$ by right-continuity. 
\item Finally, $\ironmarg(z) = g(\dist(z))$.
\end{enumerate}
\end{definition}
Step~4 of Definition~\ref{d:ironing} makes sense because~$G$
is convex function.  Convexity of~$G$ also implies that~$g$, and
hence~$\ironmarg$, is a monotone non-decreasing function.

The proof Myerson gives for ironing virtual valuations for payments
extends simply to any other kind of virtual valuation including our
virtual valuations for utility.  We summarize this in
Lemma~\ref{l:main} with a proof in Appendix~\ref{app:bayesian}.

\begin{lemma}
\label{l:main}
Let $\dist$ be a distribution function with virtual valuation
function $\marg(\cdot)$ and $\allocs(\vals)$ a monotone allocation rule.
Define $G$, $H$, and $\ironmarg$ as in Definition~\ref{d:ironing}.
Then
\begin{equation}\label{eq:char}
\expect[\vals]{\marg(\vali)\alloci(\vals)} \le
\expect[\vals]{\ironmarg(\vali)\alloci(\vals)},
\end{equation}
with equality holding if and only if
$\frac{d}{d \val} \alloci(\val) = 0$ whenever 
$G(\dist(\val)) < H(\dist(\val))$.
\end{lemma}

Our main theorem now follows easily.

\begin{theorem}
\label{t:main} 
Let $\dist$ be a distribution function with virtual valuation
function $\marg(\cdot)$.
Define $G$, $H$, and $\ironmarg$ as in Definition~\ref{d:ironing}.
For valuation profiles drawn from distribution $\dists$,
the mechanisms that maximize the expected residual
surplus are precisely those satisfying
\begin{enumerate}
\item $\allocs(\vals) \in \argmax_{\allocs' \in \feasibles}
\sum\nolimits_i \ironmarg(\vali) \alloci'$ for every $\vals$; and
\item for all $i$, 
$\frac{d}{d \val} \alloci(\val) = 0$ whenever 
$G(\dist(\val)) < H(\dist(\val))$.
\end{enumerate}
\end{theorem}

\begin{proof}
First, 
there exists a mechanism that satisfies both of the desired
properties.  To see this, consider an allocation rule that maximizes
$\sum_i \ironmarg(\vali)\alloci(\vals)$ for every $\vals$.  Such a
rule can without loss of generality be a function only of
$\ironmarg(\vali)$ and not of $\vali$ directly.
At points~$\val$ where $G(F(\val)) < H(F(\val))$, $G$ is locally linear
(since it is the convex hull of~$H$) and hence $\ironmarg(\val)$ is
locally constant.  Thus such an allocation rule will satisfy
$\frac{d}{d \val}\alloci(\val) = 0$ at all such points (for all~$i$).

A mechanism that meets both conditions simultaneously
maximizes the right-hand side of~\eqref{eq:char} while
satisfying the inequality with equality.
Lemmas~\ref{l:vv} and~\ref{l:main} imply that such a mechanism
maximizes the expected residual surplus and, conversely, that
all optimal mechanisms must meet both conditions.  
\end{proof}

Theorem~\ref{t:main} shows that maximizing the ironed virtual surplus
(for utility) is equivalent to maximizing expected residual surplus
subject to incentive-compatibility.
Different tie-breaking rules can yield different optimal mechanisms.
With symmetric participants (that is, i.i.d.\ valuations) and
a symmetric feasible region (e.g., $k$-item auctions), it is
natural to consider symmetric mechanisms, and these will play a
crucial role in our benchmark for prior-free money-burning mechanisms
(see Definition~\ref{def:opt}).

\paragraph*{Interpretation.}
%

\begin{figure}
\begin{tabular}{ccc}
 MHR & nonMHR &  antiMHR\\
(e.g., uniform) & (e.g., bimodal) & (e.g., super-exponential)\\
%
%

{\small\psset{yunit=1in,xunit=1in}
\begin{pspicture}(-.2,-.2)(1.6,1.3)

\rput[b](.4,.8){\red$\ironmarg(\val)$}

\psline[linecolor=red]{*-*}(0,.6)(1.6,.6)

\psaxes[labels=none,ticks=none]{->}(1.6,1.2)

\end{pspicture}}
&
%
%

{\small\psset{yunit=1in,xunit=1in}
\begin{pspicture}(-.2,-.2)(1.6,1.3)

\rput[b](.4,.8){\red$\ironmarg(\val)$}

\pscurve[linecolor=red]{*-*}(0,.1)(.1,.3)(.1,.3)(.6,.3)(.6,.3)(.65,.4)(.69,.55)(.69,.55)(.945,.55)(.945,.55)(1,.7)(1.2,.8)(1.3,.85)(1.6,1)

\psaxes[labels=none,ticks=none]{->}(1.6,1.2)

\end{pspicture}}
&
%
%

{\small\psset{yunit=1in,xunit=1in}
\begin{pspicture}(-.2,-.2)(1.6,1.3)

\rput[b](.4,.8){\red$\ironmarg(\val)$}

\pscurve[linecolor=red]{*-*}(0,.1)(.2,.15)(.3,.3)(.6,.4)(.69,.45)(.845,.70)(1,.8)(1.1,.85)(1.6,1)

\psaxes[labels=none,ticks=none]{->}(1.6,1.2)

\end{pspicture}}
\\
Lottery is optimal. & indirect Vickrey is optimal. &
Vickrey is optimal.
\end{tabular}
\caption{Ironed virtual residual surplus in the three cases.}
\label{fig:ironed-residual-cases}
\end{figure}
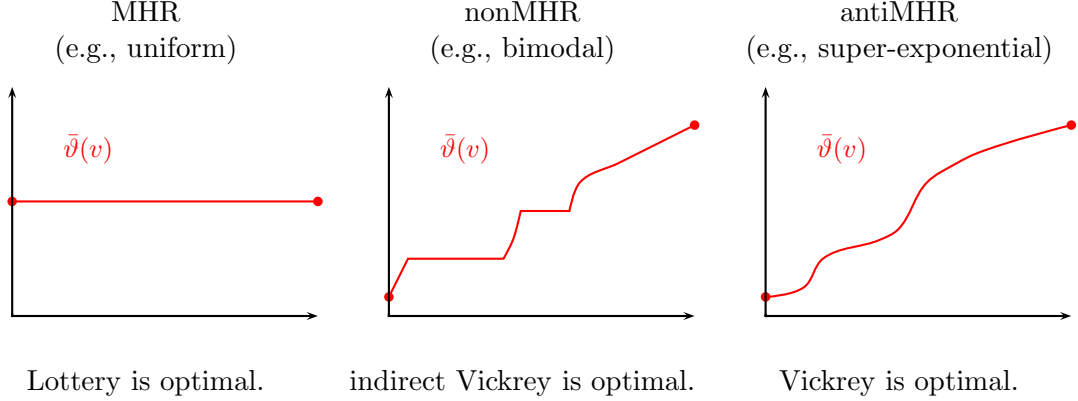


To interpret Theorem~\ref{t:main}, recall that the {\em hazard rate}
of distribution $\dist$ at $v$ is defined as
$\frac{\dens(v)}{1-\dist(v)}$.  The {\em monotone hazard rate} (MHR)
assumption is that the hazard rate is monotone non-decreasing and is a
standard assumption in mechanism design (e.g.~\cite{mye-81}).  We will
analyze this standard setting (MHR), the setting in which the hazard
rate is monotone in the opposite sense (anti-MHR), and the setting
where it is neither monotone increasing nor decreasing (non-MHR).
Notice that the hazard rate function is precisely the reciprocal
virtual valuation (for utility) function.  Our interpretation is
summarized by Figure~\ref{fig:ironed-residual-cases}.

When the valuation distribution satisfies the MHR condition,
the ironed virtual
valuations (for utility) have a special form: they are constant with
value equal to their expectation. 

\begin{lemma}
\label{l:const}
For every distribution~$\dist$ that satisfies the monotone hazard rate
condition, the 
ironed virtual valuation (for utility) function is constant with
$\ironmarg(z) = \expval$, where $\expval$ denotes the expected value
of the distribution.
\end{lemma}


\begin{proof}
Apply the ironing procedure from Definition~\ref{d:ironing} to
$\marg(z)$.  The monotone hazard rate condition implies that
$\marg(z)$ is monotone non-increasing.  Since $\dist(z)$ is monotone
non-decreasing so is $\dist^{-1}(q)$ for $q \in [0,1]$. Thus, $h(q) =
\marg(\dist^{-1}(q))$ is monotone non-increasing.  The integral $H(q)$
of the monotone non-increasing function $h(q)$ is concave.  The convex
hull $G(q)$ of the concave function $H(q)$ is a straight line.  In
particular, $H(q)$ is defined on the range $[0,1]$, so $G(q)$ is the
straight line between $(0,H(0))$ and $(1,H(1))$.  Thus, $g(q)$ is the
derivative of a straight line and is therefore constant with value
equal to the line's slope, namely~$H(1)$.  Thus, $\ironmarg(z) = H(1)$.
It remains to show that $H(1) = \expval$.  By definition,
\begin{align*}
H(1) &= \int_0^1 \marg(\dist^{-1}(q))dq.\\
\intertext{Substituting $q = \dist(z)$, $dq = \dens(z)dz$, and the support of $\dist$ as $(a,b)$, we have}
H(1) &= \int_a^b \marg(z) \dens(z) dz.\\
\intertext{Using the definition of $\marg(\cdot)$ and the definition of expectation for non-negative random variables gives}
H(1) &= \int_a^b (1 - \dist(z)) dz = \expval.
\end{align*}
\end{proof}

Therefore, under MHR the mechanism that maximizes the
ironed virtual surplus is the one that {\em maximizes the ex ante expected
surplus}, without asking for bids and without any transfers.  
For example, in a multi-unit auction with i.i.d.\ bidders, all agents
are equal ex ante, and thus any allocation rule that ignores the bids
and always allocates all $k$ units (charging nothing) is optimal.  

\begin{corollary}
\label{c:k-lottery}
For agents with i.i.d.\ valuations satisfying the MHR condition,
an optimal (symmetric) money-burning mechanism for allocating $k$
units is a $k$-unit lottery.
\end{corollary}



Suppose the distribution satisfies the anti-MHR condition
which implies that
the virtual valuation (for utility) functions are monotone
non-decreasing.  The ironed virtual valuation
function is then identical to the virtual valuation function.  The
i.i.d.~assumption implies that all agents have the same virtual
valuation function, so the agents with the highest virtual valuations
are also the agents with the highest valuations.  Therefore, an
optimal money-burning
mechanism for allocating $k$ units assigns the units to the $k$ agents
with the highest valuations.\footnote{Virtual valuations
need not be strictly increasing, so two bidders with different
valuations may have identical virtual valuations.  In the anti-MHR
case, it is permissible to break ties in favor of the
agent with the highest valuation.  In the notation of
Lemma~\ref{l:main}, $G = H$ throughout $[0,1]$, so the tie-breaking
rule does not affect the expected residual surplus.}
This is precisely the allocation rule used by the $k$-unit Vickrey auction
\cite{vic-61}, so  the truthtelling payment rule is that all
winners pay the $k+1$st highest valuation.

\begin{corollary}\label{cor:anti}
For agents with i.i.d.\ valuations satisfying the anti-MHR condition,
an optimal (symmetric) money-burning mechanism for allocating $k$
units is a $k$-unit Vickrey auction.
\end{corollary} 

To optimally allocate $k$ units of an item in the non-MHR case, we
simply award the items to the agents with the largest ironed virtual
valuations (for utility). 
Ironed virtual valuations are constant over regions in which 
non-trivial ironing takes place, resulting in potential ties among
players with distinct valuations.
The allocation rule of an optimal mechanism cannot change over
ironed regions (Lemma~\ref{l:main}), so we cannot break ties among ironed
virtual valuations in favor of agents with higher valuations.  We can
break these ties arbitrarily (e.g., based on a predetermined total
ordering of the agents) or randomly.  In either case the optimal mechanism
can be described succinctly as an {\em indirect} generalization of the
$k$-unit Vickrey auction where the bid space is restricted to be
intervals in which the ironed virtual valuation function is strictly
increasing.  The $k$ agents with the highest bids win and ties are
broken in a predetermined way.  Payments in this mechanism are given
by Lemma~\ref{l:ic} and are described in more detail for this
case in the next section.

\begin{corollary} 
For agents with i.i.d.\ non-MHR valuations,
an optimal (symmetric) money-burning
mechanism for allocating $k$ units is an indirect $k$-unit Vickrey
auction: for valuations in the range $R = [a,b]$ 
and subrange $R' \subset R$ on which $\ironmarg(\val)$ has
positive slope, it is the indirect mechanism where agents bid $\bidi
\in R'$ and the $k$ agents with the highest bids win, with
ties broken uniformly at random.
\end{corollary}

\section{Prior-Free Money-Burning Mechanism Design}\label{sec:worst}

We now depart from the Bayesian setting and design near-optimal
prior-free mechanisms for multi-unit auctions.
Section~\ref{subsec:bm} corresponds to the second step in our
prior-free mechanism design template and leverages our
characterization of Bayesian optimal mechanisms to identify a simple,
tight, and distribution-independent performance benchmark.
Section~\ref{subsec:rsol} gives a prior-free mechanism that, for every
valuation profile, obtains expected residual surplus within a constant
factor of this benchmark.  This mechanism implements the third step of
our design template.  We consider lower bounds on the approximation
ratio of all prior-free mechanisms (the final step of the template) in
Section~\ref{subsec:lb}.

For ease of discussion the payment rules we describe in this section
are for mechanism implementations that are dominant strategy incentive
compatible for agents that are risk-neutral with respect to
randomization in the mechanism, i.e., mechanisms that are {\em
truthful in expectation}.  All of these mechanisms have natural
implementations with payment rules that make them dominant strategy
incentive compatible for any fixed outcome of the mechanism's random
decisions, i.e., mechanisms that are {\em truthful all the time}.  In
the computer science literature, discussion of these distinctions can
be found in~\cite{APTT-03}.

\subsection{A Performance Benchmark for Prior-Free Mechanisms}\label{subsec:bm}

Intuitively, our performance benchmark for a valuation
profile is the maximum residual surplus achieved by a symmetric
mechanism that is optimal for some i.i.d.\ distribution.
The next definition formalizes the class of mechanisms that define the
benchmark.
\begin{definition}[$\Mye_\dists$]
\label{def:opt}
For an i.i.d.\ distribution $\dists$ with ironed virtual valuation
(for utility) function $\ironmarg$, the mechanism $\Mye_\dists$
is defined as follows.
\begin{enumerate}

\item Given $\vals$, choose a feasible allocation
maximizing $\sum_i \ironmarg(\vali) \alloci$.  If there are multiple
such allocations, choose one uniformly at random.

\item Let $\allocs$ denote the corresponding allocation rule,
with $\alloci(\vals)$ denoting the probability that player~$i$
receives an item given the valuation profile $\vals$.
Let $\prices$ denote the (unique) payment rule dictated by
Lemma~\ref{l:ic}.

\item Given valuations $\vals$ and the random choice of allocation in
the first step, charge each winner~$i$ the price 
$\pricei(\vals)/\alloci(\vals)$ and each loser~0.

\end{enumerate}
\end{definition}
By Theorem~\ref{t:main}, $\Mye_\dists$ maximizes the expected residual
surplus for valuations drawn from~$\dists$.
Using Lemma~\ref{l:ic}, it is also incentive-compatible and ex post
individually rational.
It is symmetric provided the set of feasible allocations
is symmetric (i.e., is a $k$-item auction).  In this case, the first
step awards the $k$ items to the bidders with the top~$k$ ironed
virtual valuations (for utility) with respect to the
distribution~$\dists$, breaking ties uniformly at random.

Our benchmark is then:
\begin{equation}\label{eq:bm}
\bm(\vals) = \sup\nolimits_\dists \Mye_\dists(\vals),
\end{equation}
where $\Mye_\dists(\vals)$ denotes the expected residual surplus (over
the choice of random allocation) obtained by the mechanism $\Mye_\dists$
on the valuation profile $\vals$.
This benchmark is, by definition, distribution-independent.  As such,
it provides a yardstick by which we can measure prior-free mechanisms:
we say that a (randomized) mechanism {\em $\beta$-approximates the
benchmark~$\bm$} if, for every valuation profile~$\vals$, its expected
residual surplus is at least $\bm(\vals)/\beta$.  Note the strength of
this guarantee: for example, if a mechanism
$\beta$-approximates the benchmark~$\bm$, then on any
i.i.d.~distribution it achieves at least a~$\beta$ fraction of the
expected residual surplus of every mechanism.  Naturally, no prior-free
mechanism is better than 1-approximate; we give stronger lower bounds
in Section~\ref{subsec:lb}.

\paragraph{Remark.}
Restricting attention in Definition~\ref{def:opt} to optimal
mechanisms that use symmetric tie-breaking rules is crucial for
obtaining a tractable benchmark.  
For example, when $\dists$ is an i.i.d.\ distribution satisfying the
MHR assumption, Theorem~\ref{t:main} implies that {\em every} constant
allocation rule that allocates all items (with zero payments) is optimal 
(recall Corollary~\ref{c:k-lottery}). 
For a single-item auction and a valuation profile $\vals$, say with
the first bidder having the highest valuation, the mechanism that
always awards the good to the first bidder and charges nothing
achieves the full surplus.  (Of course, this mechanism has extremely
poor performance on many other valuation profiles.)
As no incentive-compatible money-burning mechanism always achieves a
constant fraction of the full surplus (see Proposition~\ref{prop:poa_lb}), 
allowing arbitrary asymmetric optimal mechanisms to participate
in~\eqref{eq:bm} would yield an unachievable benchmark.

\subsection{Multi-Unit Auctions and Two-Price Lotteries}

The definition of~$\bm$ in~\eqref{eq:bm} is meaningful in general
single-parameter settings, but appears to be analytically tractable
only in problems with additional structure, symmetry in particular.
We next give a simple description of this benchmark, and an even
simpler approximation of it, for multi-unit auctions.

\Xcomment{
In the MHR case, the optimal mechanism is the $k$-Unit $0$-Lottery.
In the anti-MHR case, the optimal mechanism is the $k$-Unit Vickrey
Auction, which for valuation profile $\vals$ is equivalent to the
$k$-Unit $(\valith[k+1]$-Lottery
(Figure~\ref{fig:ironed-residual-cases}).  If these were the only
cases we could conclude that $\bm(\vals) = \max\{\frac{k}{n}\sum_i
\vali, \sum_{i\leq k} (\valith - \valith[k+1])$ would be appropriate.
Unfortunately, the case where the hazard rate it not monotone in
either direction requires more careful analysis.
}

What does $\Mye_\dists$ look like for such problems?  
When the distribution on valuations satisfies the
MHR assumption, $\Mye_\dists$ is a $k$-unit lottery
(cf., Corollary~\ref{c:k-lottery}).
Under the anti-MHR assumption, $\Mye_\dists$ is a $k$-unit
Vickrey auction (cf., Corollary~\ref{cor:anti}).
We can view the $k$-unit Vickrey auction, ex post, as a $k$-unit
$\valith[k+1]$-lottery,
where $\valith[k+1]$ is the $k+1$st highest valuation, in the
following sense.
\begin{definition}[$k$-unit $p$-lottery]\label{def:p-lottery}
The {\em $k$-unit $p$-lottery}, denoted $\Lottery_p$, allocates to
agents with value at least $p$ at price $p$.  If there are more than
$k$ such agents, the winning agents are selected uniformly at random.
\end{definition}

One natural
conjecture is that, ex post, the outcome of 
every mechanism of the form $\Mye_\dists$ on a
valuation profile $\vals$ looks like a $k$-unit $p$-lottery for some
value of $p$.  For non-MHR distributions~$\dists$, however,
$\Mye_\dists$ can assume the more complex form of a two-price lottery,
ex post.

\begin{definition}[$k$-unit $(p,q)$-lottery]\label{def:pq}
A {\em $k$-unit $(p,q)$-lottery}, denoted $\Lottery_{p,q}$, is the
following mechanism.  Let $s$ and $t$ denote the number of agents
with bid in the range $(p,\infty)$ and $(q,p]$, respectively.
\begin{enumerate}

\item If $s \ge k$, run a $k$-unit $p$-lottery on the top~$s$ agents.

\item If $s+t \leq k$, sell to the top $s+t$ agents at price $q$.

\item Otherwise, run a $(k-s)$-unit $q$-lottery on the agents with
bid in $(q,p]$ and allocate each of the top~$s$ agents a good at
the price dictated by Lemma~\ref{l:ic}:
$
\tfrac{k-s+1}{t+1} q + 
\tfrac{s+t-k}{t+1} p.
$
\end{enumerate}
\end{definition}

We now prove that for every i.i.d.~distribution $\dists$ and every 
valuation profile $\vals$, the mechanism $\Mye_\dists$ results in an
outcome and payments that, ex post, are identical to those of a $k$-unit
$(p,q)$-lottery.

\begin{lemma} \label{lem:benchmark} 
For every valuation profile~$\vals$, there is a $k$-unit
$(p,q)$-lottery with expected residual surplus 
$\bm(\vals)$.
%
\end{lemma}

\begin{proof}
By definition~\eqref{eq:bm}, we only need to show that, for every
i.i.d.\ distribution $\dists$ and valuation profile~$\vals$, $\Mye_\dists(\vals)$ has the same outcome as a $k$-unit $(p,q)$-lottery.

Fix~$\dists$ and~$\vals$, and assume that $\val_1 \ge \cdots \ge
\val_n$.  Thus, $\ironmarg(\val_1) \ge \cdots \ge \ironmarg(\val_n)$.
Recall by Definition~\ref{def:opt} that $\Mye_\dists$ maximizes
$\sum\nolimits_i \ironmarg(\vali)\alloci$ and breaks ties randomly.
Define $S = \{ i \,:\, \ironmarg(\val_i) > \ironmarg(\val_{k+1}) \}$,
$T = \{ i \,:\, \ironmarg(\val_i) = \ironmarg(\val_{k+1})\}$, $s =
\setsize{S}$, and $t = \setsize{T}$.  Assume we are in the more
technical case that $0 < s < k < s+t$ (the other cases
follow from similar arguments).  It is easy to see that $\Mye_\dists$
assigns a unit to each bidder in~$S$ and allocates the remaining $k-s$
units randomly to bidders in~$T$.  Let $q = \inf \{ \val \,:\,
\ironmarg(\val) = \ironmarg(\val_{k+1}) \}$ and 
$p = \inf \{ \val \,:\, \ironmarg(\val) > \ironmarg(\val_{k+1}) \}$. 
The allocation is thus identical to a $k$-unit $(p,q)$-lottery.  It
remains to show that the payments are correct.

Let $\alloci(\cdot)$ be as in Definition~\ref{def:opt}.
Consider agent $i \in T$.  If $i$ bids below $q$ then $i$ loses, while
if $i$ bids at least $q$ then $i$ wins with the same probability as
when $i$ bids $\vali$.  Therefore, $\alloci(\val)$ for $\val \leq \vali$
is step function at $\val = q$.  Thus, $\pricei(\vali) = \vali
\alloci(\vali) - \int_0^{\vali} \alloci(\val) d\val = q
\alloci(\vali)$ and $i$'s payment on winning is 
$\pricei(\vali) / \alloci(\vali) = q$, as in
the $k$-unit $(p,q)$-lottery.
Now consider an agent $i \in S$.  If $i$ were to bid $\val < q$, $i$ would
lose, i.e., $\alloci(\val) = 0$.  If $i$ were to bid $\val \in [q,p)$
then $i$ would leave the set $S$ of agents guaranteed a unit, and
would join the set $T$, making $t+1$ agents who would share
$s-k+1$ remaining items by lottery.  In this case, $\alloci(\val) =
\frac{s-k+1}{t+1}$.  Of course, $\alloci(\val) = 1$
when $\val > p$.
As $\alloci(\cdot)$ is identical to the allocation function for agent
$i$ in the $k$-unit $(p,q)$-lottery, the payments are also identical.
\end{proof}

As we have seen, mechanisms of the form $\Mye_\dists$ can produce
outcomes not equivalent to that of a single-price lottery.
Our next lemma
shows that $k$-unit $p$-lotteries give 2-approximations to $k$-unit
$(p,q)$-lotteries.
This allows us to relate the performance of single-price
lotteries to our benchmark (Corollary~\ref{cor:lottery}),
which will be useful in our construction of an approximately optimal
prior-free mechanism in the next section. 

\begin{lemma} \label{lem:lotteries}
For every valuation profile $\vals$ and parameters $k$, $p$, and $q$,
there is a $p'$ such that the $k$-unit $p'$-lottery obtains at least
half of the expected residual surplus of the $k$-unit
$(p,q)$-lottery.
\end{lemma}

\begin{proof}
We prove the lemma by showing that $\Lottery_{p,q}(\vals) \le
\Lottery_{p}(\vals) + \Lottery_{q}(\vals)$.
We argue the stronger statement that
each agent enjoys at least as large a combined expected
utility in $\Lottery_{p}(\vals)$ and $\Lottery_{q}(\vals)$ as in
$\Lottery_{p,q}(\vals)$.

Let $S$ and $T$ denote the agents with values in the ranges
$(p,\infty)$ and $(q,p]$, respectively.  Let $s = \setsize{S}$ and $t
= \setsize{T}$.  Assume that $0 < s < k < s+t$ as otherwise the
$k$-unit $(p,q)$ lottery is a single-price lottery.  Each agent in~$T$
participates in a $k$-unit $q$-lottery in $\Lottery_{q}$ and only a
$(k-s)$-unit $q$-lottery in $\Lottery_{p,q}$; its expected utility can
only be smaller in the second case.  Now consider $i \in S$.  Writing
$r = (k-s+1)/(t+1)$, we can upper bound the utility of an agent $i$ in $\Lottery_{p,q}$ by
$$v_i - rq - (1-r)p = 
(1-r)(v_i-p) + r(v_i-q)
\le (v_i-p) + \tfrac{k}{s+t} \cdot (v_i-q),$$
which is the combined expected utility that the agent obtains from
participating in both a $k$-unit $p$-lottery (with $s < k$) and a
$k$-unit $q$-lottery.
\end{proof}

\begin{corollary} \label{lem:lottery} \label{cor:lottery} 
For every valuation profile~$\vals$, there is a $k$-unit $p$-lottery
with expected residual surplus at least $\bm(\vals)/2$.
\end{corollary}

\subsection{A Near-Optimal Prior-Free Money-Burning Mechanism}
\label{subsec:rsol}

\newcommand{\sm}{{\setminus}}
\newcommand{\sse}{{\subseteq}}
\newcommand{\event}{{\cal E}}
\newcommand{\lb}{{ ??}}

We now give a prior-free mechanism that
$O(1)$-approximates the benchmark~$\bm$.  This mechanism is motivated
by the following observations.  First, by Corollary~\ref{cor:lottery},
our mechanism only needs to compete with $k$-unit $p$-lotteries.  Second, if
many agents make significant contributions to the optimal residual
surplus, then we can use random sampling techniques to approximate the
optimal $k$-unit $p$-lottery.  Third, if a few agents are single-handedly
responsible for the residual surplus obtained by the optimal
$k$-unit $p$-lottery, then the $k$-unit Vickrey auction obtains a 
constant fraction of the optimal residual surplus.
The precise mechanism is as follows.

\begin{definition}[Random Sampling Optimal Lottery ($\RSOL$)]
With a set $S=\{1,\ldots,n\}$ of $n$ agents and a supply of $k$
identical units of an item, the {\em Random Sampling Optimal Lottery
($\RSOL$)} is the following mechanism.
\begin{enumerate}

\item Choose a subset $S_1 \subset S$ of the agents uniformly at random, and
let $S_2$ denote the rest of the agents.
Let $p_2$ denote the price charged by the optimal $k$-unit $p$-lottery
for $S_2$.

\item With 50\% probability,
run a $k$-unit $p_2$-lottery on~$S_1$.

\item Otherwise, run a $k$-unit Vickrey auction on~$S_1$.

\end{enumerate}
\end{definition}

We have deliberately avoided optimizing this mechanism in order to keep
its description and analysis 
as simple as possible.

\begin{theorem}\label{thm:worst}
$\RSOL$ $O(1)$-approximates the benchmark $\bm$.
\end{theorem}

In our proof of Theorem~\ref{thm:worst}, we use the following
``Balanced Sampling Lemma'' of 
Feige et al.~\cite{FFHK-05} to control the similarity between the
random sample $S_1$ chosen by $\RSOL$ and its complement~$S_2$.

\begin{lemma}[Balanced Sampling Lemma~\cite{FFHK-05}]
Let $S$ be a random subset of $\{1,2,\ldots,n\}$.  Let $n_i$ denote
$|S \cap \{1,2,\ldots,i\}|$.
Then
$$
\prob{n_i \le \tfrac{3}{4}i~\text{for all $i \in \{1,2,\ldots,n\}$} 
\,\big|\, n_1 = 0} \ge \tfrac{9}{10}.
$$
\end{lemma}

\begin{proof} (of Theorem~\ref{thm:worst}).
Fix a valuation profile~$\vals$ with $\val_1 \ge \cdots \ge \val_n$
and a supply~$k \ge 1$.
For clarity, we make no attempt
to optimize the constants in the following analysis.

We analyze the performance of $\RSOL$ only when 
certain sampling events occur.
For $i=1,2$, let $\event_i$ denote
the event that agent~$i$ is included in the set~$S_i$.  
Clearly, $\prob{\event_1 \cap \event_2} = 1/4$.  Conditioning on
$\event_1 \cap \event_2$, 
let $\event_3$ denote the event that the
Balanced Sampling Lemma holds for the sample $S_1 \sm \{1\}$ when
viewed as a subset of $\{2,3,\ldots,n\}$.
Similarly, let $\event_4$ denote the event that the
Balanced Sampling Lemma holds for the sample $S_2 \sm \{2\}$ when
viewed as a subset of $\{1,3,\ldots,n\}$.
By the Principle of Deferred Decisions and the Union Bound,
$\prob{\event_3 \cap \event_4 | \event_1 \cap 
\event_2} \ge 4/5$.  Hence, $\prob{\cap_{i=1}^4 \event_i} \ge 1/5$.
We prove a bound on the approximation ratio conditioned 
on the event $\cap_{i=1}^4 \event_i$; since the mechanism always has
nonnegative residual surplus, its unconditional approximation ratio is
at most~5 times as large.

Let $n_i$ and $\bar{n}_i$ denote  $|S_1 \cap \{1,2,\ldots,i\}|$ and
$|S_2 \cap \{1,2,\ldots,i\}|$, respectively.  Since the event
$\cap_{i=1}^4 \event_i$ holds, we have 
\begin{equation}\label{eq:balance}
n_i,\bar{n}_i \in \left[ \tfrac{1}{6}i, \tfrac{5}{6}i \right]
\end{equation}
for every $i \in \{2,3,\ldots,n\}$, and also $n_1 = 1$ and $\bar{n}_1
= 0$.

By Corollary~\ref{cor:lottery}, we only need to
show that the expected residual surplus of the mechanism is at least a
constant fraction of that of the optimal $k$-unit $p$-lottery for $\vals$. 
For a subset~$T$ of agents and a price~$p$, let $W(T,p)$ denote the
residual surplus of the $k$-unit $p$-lottery for~$T$.  Letting
$n^T_i$ denote $|T \cap \{1,2,\ldots,i\}|$ and
$d_i$ denote $\val_i - \val_{i+1}$ for $i \in \{1,2,\ldots,n\}$
(interpreting $\val_{n+1} = 0$), for every $\ell$ we obtain the
following useful identity:
\begin{equation}\label{eq:W}
W(T,\val_{\ell+1}) = 
\frac{\min\{k,n^T_{\ell}\}}{n^T_{\ell}} \left(\sum_{i \in T \cap
\{1,\ldots,\ell\}} \val_i
\right) -
\min\{k,n^T_{\ell}\} \cdot \val_{\ell+1} = 
\frac{\min\{k,n^T_{\ell}\}}{n^T_{\ell}} \sum_{i=1}^{\ell} n^T_id_i.
\end{equation}

Let $v_{\ell^*+1}$ denote the optimal price for a $k$-unit $p$-lottery for
$\vals$, and note that $\ell^* \ge k$.
By~\eqref{eq:W}, the residual surplus of this optimal lottery is
$$
W(S,\val_{\ell^*+1}) = \frac{k}{\ell^*} \sum_{i=1}^{\ell} id_i.
$$
To analyze the expected residual surplus of $\RSOL$, first
suppose that it executes a $k$-unit $p_2$-lottery where~$p_2 =
\val_{m+1}$ for some $m$.
We then have
$$
W(S_2,p_2)  \ge  W(S_2,\val_{\ell^*+1})
 =  
\frac{\min\{k,\bar{n}_{\ell^*}\}}{\bar{n}_{\ell^*}} \sum_{i=1}^{\ell^*}
\bar{n}_id_i
 \ge 
\frac{k}{\ell^*} \sum_{i=2}^{\ell^*} \frac{i}{6}d_i
\ge \frac{W(S,\val_{\ell^*+1})}{6} - d_1,
$$
where the first inequality follows from the optimality of~$p_2$
for~$S_2$, the first equality follows from~\eqref{eq:W}, and the
second inequality follows from~\eqref{eq:balance}.  On the other hand,
inequality~\eqref{eq:balance} and a similar derivation shows that the
price~$p_2$ is nearly as effective for~$S_1$:
$$
W(S_1,p_2)  =  
\frac{\min\{k,n_{m}\}}{n_m} \sum_{i=1}^{m} n_id_i
 \ge 
\left(\frac{1}{5} \cdot \frac{\min\{k,\bar{n}_{m}\}}{\bar{n}_m} \right)
\sum_{i=1}^{m} \frac{\bar{n}_i}{5}d_i
= \frac{W(S_2,p_2)}{25} \ge \frac{W(S,\val_{\ell^*+1})}{150} - d_1.
$$

Finally, if the mechanism executes a $k$-unit Vickrey auction for~$S_1$,
then it obtains residual surplus at least $v_1 - v_2 = d_1$
(since the first agent is in~$S_1$).  Averaging the residual surplus
from the two cases proves that $\RSOL$ $O(1)$-approximates
$\bm$.
\end{proof}

\vspace{.1in}

We can improve the approximation factor in Theorem~\ref{thm:worst} by
more than an order of magnitude by modifying $\RSOL$ and
optimizing the proof.  Obtaining an approximation factor less than~10,
say, appears to require a different approach.

\section{Lower Bounds for Prior-Free Money-Burning Mechanisms}\label{subsec:lb}

This section establishes a lower bound of~$4/3$ on the approximation ratio of
every prior-free money-burning mechanism.
This implements the
fourth step of the prior-free mechanism design template outlined in
the Introduction.  Our proof follows from showing that there is a
i.i.d.~distribution~$\dists$ for which the expected value of our
benchmark~$\bm$ is a constant factor larger than the expected
residual surplus of an optimal mechanism for the distribution, such as
$\Mye_{\dists}$.  This shows an inherent gap in the prior-free
analysis framework that will manifest itself in the approximation
factor of every prior-free mechanism. 

\begin{proposition}
No prior-free money-burning mechanism has approximation ratio better
than $4/3$ with respect to the benchmark~$\bm$, even for the special
case of two agents and one unit of an item.
\end{proposition}

\begin{proof}
Our plan to exhibit a distribution over valuations such that the
expected residual surplus of the Bayesian optimal mechanism is at most
$3/4$ times that of the expected value of the benchmark~$\bm$.  It
follows that, for every randomized mechanism, there exists a valuation
profile $\vals$ for which its expected residual surplus is at most
$3/4$ times $\bm(\vals)$.

Suppose there are two agents with valuations drawn i.i.d.\ from a
standard exponential distribution with density $f(x) = e^{-x}$ on
$[0,\infty)$.  There is a single unit of an item.
This distribution has constant hazard rate, so
a lottery is an optimal mechanism (as is every mechanism that always
allocates the item and charges payments according to
Lemma~\ref{l:ic}).
The expected (residual) surplus of this mechanism is~1.

To calculate the expected value of~$\bm(\vals)$, first note that for a
valuation profile $(\val_1,\val_2)$ with $\val_1 \ge \val_2$,
the optimal $(p,q)$-lottery either chooses
$p = q = 0$ or $p = \val_2$ and $q = 0$.  Thus,
$$
\bm(\vals) = \max \left\{ \tfrac{\val_1+\val_2}{2}, \val_1 -
\tfrac{\val_2}{2} \right\}.
$$
Next, note that $(\val_1+\val_2)/2 \ge \val_1-(\val_2/2)$
if and only if $\val_1 \le 2\val_2$.

Now condition on the smaller valuation $v_2$ and write $v_1 = v_2 + x$
for $x \ge 0$. 
Since the exponential distribution is memoryless, $x$ is exponentially
distributed.  
Thus, $\expect{\bm(\val_1,\val_2) | \val_2}$ can be computed as follows (integrating over
possible values for $x \in [0,\infty)$):
\begin{eqnarray*}
\expect{\bm(v_1,v_2) | v_2}
& = &
\int_0^{v_2} \left( v_2+\frac{x}{2} \right) e^{-x}dx
+ 
\int_{v_2}^{\infty} \left(\frac{v_2}{2} + x\right)e^{-x}dx\\
& = &
v_2(1-e^{-v_2}) + \tfrac{1}{2} \left( 1 - (v_2 + 1)e^{-v_2} \right)
+ \tfrac{v_2}{2}e^{-v_2} + (v_2 + 1)e^{-v_2}\\
& = &
v_2 + \tfrac{1}{2} \left( 1 + e^{-v_2} \right).
\end{eqnarray*}

The smaller value~$v_2$ is distributed according to an exponential
distribution with rate~2.
Integrating out yields
\begin{eqnarray*}
\expect{\bm(v_1,v_2)}
& = &
\int_{0}^{\infty} (2e^{-2x})\left(x + \tfrac{1}{2} + \tfrac{1}{2}e^{-x}
\right)dx\\
& = &
\tfrac{1}{2} + \tfrac{1}{2} + \int_0^{\infty} e^{-3x}dx\\
& = & \tfrac{4}{3}.
\end{eqnarray*}
\end{proof}

For the special case of two agents and a single good, an appropriate
mixture of a lottery and the Vickrey auction is a $3/2$-approximation
of the benchmark~$\bm(\vals)$.  Determining the best-possible
approximation ratio is an open question, even in the two agent, one
unit special case.

\begin{proposition}
For two bidders and a single unit of an item, there is a prior-free
mechanism that $3/2$-approximates the benchmark~$\bm$.
\end{proposition}

\begin{proof}
Consider a valuation profile with $\val_1 \ge \val_2$.
If we run a Vickrey auction with probability~$1/3$ and a lottery with
probability~$2/3$, then the expected residual surplus is
$$
\tfrac{1}{3}\left( v_1 - v_2 \right)
+ \tfrac{2}{3} \left( \tfrac{v_1+v_2}{2} \right)
= \tfrac{2}{3}v_1
\ge \tfrac{2}{3} 
\max \left\{ \tfrac{\val_1+\val_2}{2}, \val_1 -
\tfrac{\val_2}{2} \right\} = \tfrac{2}{3}
\bm(\vals).
$$
\end{proof}

\section{Quantifying the Power of Transfers and Money-Burning}\label{sec:poa}

For the objective of surplus maximization, mechanisms with general
transfers are clearly as powerful as money-burning mechanisms, which
in turn are as powerful as mechanisms without money.  This section
quantifies the distance between the levels of this hierarchy by
studying surplus approximation in multi-unit auctions.  Precisely, we
call a class of mechanisms {\em
$\alpha$-surplus maximizers} if, for every multi-unit auction problem,
there is a mechanism in the class that obtains at least a $1/\alpha$
fraction of the full surplus for every valuation profile.  For
example, mechanisms with transfers are 1-surplus maximizers, because
the VCG mechanism achieves full surplus in every multi-unit auction
problem.  Mechanisms without transfers are $(n/k)$-surplus maximizers,
since the expected surplus of a $k$-unit lottery is $k/n$ times the
full surplus.  One can show (details omitted) that mechanisms without
transfers are not significantly better than $\Theta(n/k)$-surplus
maximizers.

The interesting question is to identify the exact location of
money-burning mechanisms between these two extremes: what is the
potential benefit of implementing monetary transfers in a system that
initially only supports money burning?
We give a lower bound 
and a matching upper bound, for all $k$ and $n$.

\begin{proposition}\label{prop:poa_lb}
Money-burning mechanisms are $\Omega(1 + \log \tfrac{n}{k})$-surplus
maximizers in $k$-unit auctions.
\end{proposition}

\begin{proof}
By Yao's Minimax Theorem, we only need to lower bound the surplus
approximation achieved by an optimal mechanism on a worst-case
distribution over valuation profiles.

Fix $k$ and draw $n$ valuations i.i.d.\ from an exponential
distribution (with density $e^{-x}$ on $[0,\infty)$).  This
distribution has constant hazard rate and so, by our results in
Section~\ref{sec:bayesian}, the $k$-unit lottery maximizes
the expected residual surplus.  Since the expected valuation of every
bidder is~1, the expected (residual) surplus of this mechanism is~$k$.

The expected value of the full surplus is that of the sum of the top~$k$
out of~$n$ i.i.d.\ samples of an exponential distribution.  A
calculation shows that this expectation equals~$\Theta(k(1 +  \log
\tfrac{n}{k}))$, completing the proof.
\end{proof}

\begin{theorem}\label{thm:ub}
Money-burning mechanisms are $O(1 + \log \tfrac{n}{k})$-surplus
maximizers in $k$-unit auctions.
\end{theorem}

\begin{proof}
Fix $k$ and a valuation profile~$\vals$ with $\val_1 \ge \cdots \ge \val_n$.
Assume for simplicity that both $k$ and $n$ are powers of~$2$.
Our simple mechanism is as follows.  First, 
choose a nonnegative integer~$j$ uniformly at random, subject to $k
\le 2^j \le n$.  Note that there are $1+ \log_2 (n/k)$ possible choices
for~$j$.  Second, run a $k$-unit $\val_{2^j+1}$-lottery, where we interpret
$\val_{n+1}$ as zero.

Write $V^* = \sum_{i=1}^k \val_i$ for the full surplus.
For $j \in \{ \log_2 k, \ldots, \log_2 n \}$, let $R_j$ denote the
residual surplus obtained by the mechanism for a given value of~$j$.
We claim that
$$
\expect{R_j \,|\, \mbox{$j$ is chosen}} \ge 
\left\{ \begin{array}{lr} 
\tfrac{V^*}{2} - \tfrac{k}{2}v_{k+1} & \mbox{if $j = \log_2 k$}\\
\tfrac{k}{2} \left( v_{2^{j-1}+1} - v_{2^j+1} \right) & \mbox{otherwise}.
\end{array} \right. $$
When $j = \log_2 k$, the residual surplus is exactly
$V^* - kv_{k+1} \ge (V^* - kv_{k+1})/2$.
To justify the second case, note that~$k$ units will be randomly
allocated amongst the top $2^j$ bidders at price $v_{2^j+1}$.
Each of these goods is allocated to one of the top $2^{j-1}$ of these
bidders with 50\% probability, and the residual surplus contributed by
such an allocation is at least 
$v_{2^{j-1}} - v_{2^j+1} \ge v_{2^{j-1}+1} - v_{2^j+1}$.

Let~$R$ denote the residual surplus obtained by our mechanism.  The
following derivation completes the proof:
\begin{eqnarray*}
\expect{R} 
& = & \sum\nolimits_{j=\log_2 k}^{\log_2 n} 
\expect{R_j \,|\, \mbox{$j$ is chosen}} 
\cdot \prob{\mbox{$j$ is chosen}}\\
& \ge & \tfrac{1}{1 + \log_2 (n/k)} 
\left( \tfrac{V^*}{2} - \tfrac{k}{2}v_{k+1} + 
\sum\nolimits_{j=1 + \log_2 k}^{\log_2 n} 
\tfrac{k}{2} \left( v_{2^{j-1}+1} - v_{2^j+1} \right)
\right)\\
& = & \tfrac{V^*}{2(1 + \log_2 (n/k))}.
\end{eqnarray*}
\end{proof}

Since the mechanism in Theorem~\ref{thm:ub} is prior-free, we obtain
the same (tight) guarantee for every Bayesian optimal mechanism.

\begin{corollary}\label{cor:poa_bayes}
For every i.i.d.~distribution~$\dists$, the expected residual surplus of the
Bayesian optimal mechanism for~$\dists$ obtains an $\Omega(1/(1 +
\log(n/k)))$ fraction of the expected full surplus.
\end{corollary}

Theorem~\ref{thm:ub} and Corollary~\ref{cor:poa_bayes}
suggest that the cost of implementing money-burning
payments instead of (possibly expensive or infeasible) general
transfers is relatively modest, provided an optimal money-burning
mechanism is used.

\section{Conclusions}\label{sec:conc}

We phrased our analysis of the Bayesian setting in terms of
feasible allocations 
(e.g., $\allocs \in \feasibles$ if and only if $\sum_i \alloci \leq k$
for the $k$-unit auction problem); however, it applies more generally to
single-parameter agent problems where the service provider must
pay an arbitrary cost $\cost(\allocs)$ for the allocation $\allocs$
produced.  Standard problems in this setting include fixed cost
services, non-excludable public goods, and multicast
auctions~\cite{FPS-00}.  The solution to these problems is again to
maximize the ironed virtual surplus, which in this context is the sum
of the agents' ironed virtual valuations less the cost of providing
the service, $\sum_i \ironmargi(\vali)x_i - \cost(\allocs)$.  This
generalization also applies when the agents'
valuations are independent but 
not identically distributed, i.e., agent $i$ has ironed
virtual valuation function $\ironmargi(\cdot)$.

\begin{theorem}
\label{t:main-costs} 
Given service cost $\cost(\cdot)$ and a valuation profile, $\vals$, drawn from
distribution $\dists = \disti[1] \times \cdots \times \disti[n]$ with
ironed virtual valuation (for utility) function $\ironmargi(\cdot)$
for agent $i$, every mechanism with allocation rule satisfying
\begin{enumerate}
\item $\allocs(\vals) \in \argmax_{\allocs'} \sum\nolimits_i
\ironmargi(\vali) \alloci - \cost(\allocs')$ and
\item 
$\tfrac{d}{d \vali} \ironmargi(\vali) = 0
\Rightarrow
\tfrac{d}{d \vali} \alloci(\vali) = 0$
\end{enumerate}
is optimal with respect to expected residual surplus.
\end{theorem}

Our results for the Bayesian problem also extend beyond dominant
strategy mechanisms.  The well known {\em revenue equivalence} result
\cite{mye-81} is popularly stated as: first price, second price
(a.k.a., Vickrey), and all-pay auctions all achieve the same profit.
Of course this applies to money burning as well.  While this paper
emphasized the dominant strategy ``second price'' optimal auction,
there are also first-price and all-pay variants that achieve the same
performance.  The all-pay variant is especially interesting because of
its potential usefulness for network problems. For example, in network
routing, all agents could attach a proof of a computational
payment to their packets.  The routing protocol can then route the
appropriate packets (depending on the amount of computational
payment) and drop the rest.  There is no need for a round of
bidding, a round of transmitting the packets of winning agents, and a
round of collecting payments.

One of our main results is in giving a benchmark based on the optimal
mechanism for the symmetric setting of i.i.d.~agents and $k$-unit
auctions.  Another main result is in approximating this benchmark with
a prior-free mechanism.  Can these techniques be generalized beyond
symmetric settings?  In particular, the notion that agents' private
valuations may be paired with publicly observable {\em attributes}
allowed for prior-free mechanisms to approximate Bayesian mechanisms
for digital good auctions and non-identically distributed valuations
\cite{BBHM-05}.  Further, there has been some limited success in
prior-free optimal mechanism design with structured costs or feasible
allocations (e.g., \cite{FGHK-02} for multicast auctions and
\cite{KKT-05} for path auctions).

Our analyses and the prior-free template extend to $k$-unit auction
problems beyond our objective of residual surplus.  Imagine the
$k$-unit auction in an i.i.d.~Bayesian setting where the optimal
solution is characterized by optimizing an ironed virtual value for
some quantity other than utility.  For example, the ``virtual
valuation for a 8\% government sales tax'', to optimize the value of
the agents and mechanism less the tax deducted by government, would be
$\virt(\val) = 0.92 \val - 0.08 \tfrac{1-\dist(\val)}{\dens(\val)}$.
The optimal $k$-unit $(p,q)$-lottery is still the appropriate
benchmark.  Furthermore, as long as the optimal $(p,q)$-lottery makes
use of prices $p,q$ bounded above by the second highest bid,
$\valith[2]$, 
as in the money-burning context, then it is likely that our prior-free
mechanism, RSOL, can be employed to approximate the benchmark.  Notice
that when applying this technique to ``virtual valuations for
payments'', which are the appropriate notion of virtual valuations for
the objective of profit maximization, the optimal $k$-unit
$(p,q)$-lottery is simply a posted price at $p$.  Furthermore, the
optimal posted price might satisfy $p = \valith[1] \gg \valith[2]$.
As it is not possible to approximate such a benchmark to within any
constant factor, the prior-free digital goods auction literature
excludes this possibility by defining the benchmark to be the profit
of the optimal posted price $p \leq \valith[2]$.

In our work there was an implicit, publicly known, exchange rate for
money burnt. In network settings, where burnt payments
correspond to degraded service quality or computational payments,
the designer may not know each agent's relative disutility for 
such payments.
This motivates considering the more general setting where agents have
a private value for burnt money in addition to their private value for
service.  This moves the problem from a single-parameter setting to
the much more challenging multi-parameter setting where optimal
mechanism design has very few positive results.

There are additionally a few loose ends to tie up with the particular
question of money-burning. For $k$-unit auctions, can we give tighter
upper and lower bounds for prior-free money-burning mechanisms with a
small number of agents?  For general settings beyond
i.i.d.~distributions and $k$-unit auctions, can we quantify the power
of money burning?

\Xcomment{
\begin{itemize}

\item simple approximation of benchmark beyond multi-unit auctions

\item constant-factor prior-free approximations beyond multi-unit
auctions

\item cost of money-burning beyond multi-unit auctions

\item for starters: how about matroid domains?

\item further applications of our prior-free template

\item tight prior-free bounds for a small number of bidders

\end{itemize}
}

\bibliographystyle{plain}
\bibliography{auctions}

\appendix

\section{Proof of Lemma~\ref{l:main}}\label{app:bayesian}

Our proof of Lemma~\ref{l:main} is based on the following lemma.

\begin{lemma} 
\label{l:decomp} For every monotone allocation rule $\alloci(\vals)$,
$$ 
\expect[\vals]{\marg(\vali) \alloci(\vals)} =
\expect[\vals]{\ironmarg(\vali) \alloci(\vals)} - \int_a^b
\left[H(\dist(\vali)) - G(\dist(\vali))\right]
\alloci'(\vali)d\vali.
$$
\end{lemma}


\begin{proof} 
Recall that $\alloci(\vali)$ is the probability of allocating to agent
$i$ with their value is $\vali$ and other agents' values are
distributed according to $\dists$: $\alloci(\vali) =
\expect[\valsmi]{\alloci(\vali,\valsmi)}.$ We use $\alloci'(\vali)$ to
denote the derivative of $\alloci(\vali)$ with respect to $\vali$.

By the definition of~$g$ and~$h$ in Definition~\ref{d:ironing},
$\marg(\vali) = \ironmarg(\vali) + h(\dist(\vali)) - g(\dist(\vali))$ 
for every~$\vali$.  Hence,
\begin{align}
\label{eq:first}
\expect[\vals]{\marg(\vali) \alloci(\vals)}
&= \expect[\vals]{\ironmarg(\vali) \alloci(\vals)}
+ \expect[\vals]{\left(h(\dist(\vali)) - g(\dist(\vali))\right)\alloci(\vals)}.\\
\intertext{Since~$\dists$ is a product distribution, the second term satisfies}
\notag
\expect[\vals]{(h(\dist(\vali)) - g(\dist(\vali)))\alloci(\vals)}
&= \int_\vals \left(h(\dist(\vali)) 
    - g(\dist(\vali))\right)\alloci(\vals)\dens(\vals)d\vals\\
&= \int_a^b \left(h(\dist(\vali)) 
    - g(\dist(\vali))\right)\alloci(\vali)\dens(\vali)d\vali.\\
\intertext{Now, integrate by parts to obtain}
\notag
\expect[\vals]{(h(\dist(\vali)) - g(\dist(\vali)))\alloci(\vals)}
&= \left[ H(\dist(\vali)) - G(\dist(\vali))\right]\alloci(\vali) \Big|_a^b
   - \int_a^b \left[H(\dist(\vali)) - G(\dist(\vali))\right]
   \alloci'(\vali)d\vali\\
\label{eq:second}
&= -\int_a^b \left[H(\dist(\vali)) - G(\dist(\vali))\right] \alloci'(\vali)d\vali.
\end{align}
Equation \eqref{eq:second} follows from the fact that, as the convex hull of $H(\cdot)$ on interval $(0,1)$, $G(\cdot)$ satisfies $G(0) = H(0)$ and $G(1) = H(1)$. Combining this with equation~\eqref{eq:first} gives the lemma.
\end{proof}

\vspace{.1in}

Now we restate and prove our main technical lemma for Bayesian optimal 
money-burning mechanisms.

\paragraph*{Lemma~\ref{l:main}}
{\em 
Let $\dist$ be a distribution function with virtual valuation
function $\marg(\cdot)$ and $\allocs(\vals)$ a monotone allocation rule.
Define $G$, $H$, and $\ironmarg$ as in Definition~\ref{d:ironing}.
Then
$$
\expect[\vals]{\marg(\vali)\alloci(\vals)} \le
\expect[\vals]{\ironmarg(\vali)\alloci(\vals)},
$$
with equality holding if and only if
$\frac{d}{d \val} \alloci(\val) = 0$ whenever 
$G(\dist(\val)) < H(\dist(\val))$.
}

\vspace{.2in}

\begin{proof}
Again, let $\alloci'(\val) = \tfrac{d}{d \val} \alloci(\val)$ be the
derivative of $\alloci(\val)$.  From Lemma~\ref{l:decomp},
\begin{equation}\label{eq:integral}
\expect[\vals]{\marg(\vali)\alloci(\vals)} = 
\expect[\vals]{\ironmarg(\vali)\alloci(\vals)}
- \int_a^b \left[H(\dist(\vali)) - G(\dist(\vali))\right] \alloci'(\vali)d\vali.
\end{equation}
Since~$G$ is the convex hull of~$H$, $G \le H$ on $[a,b]$.
Since~$x$ is a monotone allocation rule, its derivative is nonnegative.
The integral on the right-hand side of~\eqref{eq:integral} is
therefore nonnegative.  If $\alloci'(\vali) \neq 0$ only when
$G(F(\vali)) = H(F(\vali))$, then the integral vanishes.  Conversely,
since $G$ and $H$ (and hence $H-G$) are continuous, if
$\alloci'(\vali) > 0$ at a point where $G(F(\vali)) < H(F(\vali))$,
then the integral is strictly positive.
\end{proof}

\end{document}